\documentclass{article}
\usepackage{times,amsmath,amsfonts,epsfig,amssymb,amsfonts}

\def\squareforqed{\hbox{\rlap{$\sqcap$}$\sqcup$}}
\def\qed{\ifmmode\squareforqed\else{\unskip\nobreak\hfil
\penalty50\hskip1em\null\nobreak\hfil\squareforqed
\parfillskip=0pt\finalhyphendemerits=0\endgraf}\fi}

\newtheorem{theorem}{Theorem}

\newtheorem{lemma}[theorem]{Lemma}

\newtheorem{example}[theorem]{Example}

\newtheorem{definition}[theorem]{Definition}

\newenvironment{proof}{\begin{trivlist}\item[]{\flushleft\bf Proof }}
{\qed\end{trivlist}}

\newcommand{\ket}[1]{| #1 \rangle}
\newcommand{\bra}[1]{\mbox{$\langle #1 |$}}
\newcommand{\inner}[2]{\mbox{$\langle #1 | #2 \rangle$}}
\newcommand{\tr}[1]{\mbox{Tr} \, #1 }
\newcommand{\vis}{\text{v}}

\DeclareMathSymbol{\leqslant}{\mathrel}{AMSa}{"36}

\begin{document}

\title{On the Complexity of Quantum Languages}

\author{
Elham Kashefi\\
{\protect\small\sl Computing Laboratory, Oxford University\/}%
\thanks{\,Email: \texttt{elham.kashefi}\texttt{$\mathchar"40$comlab.ox.ac.uk}.}
\and
Carolina Moura Alves\\
{\protect\small\sl Clarendon Laboratory, Oxford University\/}%
\thanks{\,Email: \texttt{carolina.mouraalves}\texttt{$\mathchar"40$physics.ox.ac.uk}.}}

\maketitle
\date{}


\begin{abstract}

The standard inputs given to a quantum machine are classical
binary strings. In this view, any quantum complexity class is a
collection of subsets of $\{0,1\}^{*}$. However, a quantum machine
can also accept quantum states as its input. T. Yamakami has
introduced a general framework for quantum operators and inputs
\cite{Yam02}. In this paper we present several quantum languages
within this model and by generalizing the complexity classes QMA
and QCMA we analyze the complexity of the introduced languages. We
also discuss how to derive a classical language from a given
quantum language and as a result we introduce new QCMA and QMA
languages.

\end{abstract}


\section{Introduction}

One of the goals of complexity theory is to classify problems as
to their intrinsic computational complexity. To date researchers
have made a great deal of progress in classifying {\em classical
problems} into general complexity classes, which characterize at
least in a rough way their inherent difficulty. While the
definition of classical complexity classes is based on a classical
model, the definition of quantum complexity classes is based on a
quantum machine.

Loosely speaking, a classical problem is a relation of strings
over the alphabet $\{0,1\}$. Accordingly, the inputs to a
classical or quantum machine are classical (binary) strings which
represent {\em instances} of the underlying problem. A quantum
machine can also accept quantum states as its input. In this
general picture we consider a {\em quantum problem} to be a
property of quantum states, that is checkable by a quantum
machine. Within this paradigm we aim to decide whether a quantum
state satisfies a given property or whether it is far from all
quantum states satisfying that property. This can be considered as
an extension of the quantum property testing where one uses a
quantum machine to test a property of classical objects
\cite{FMSS03,BFNR03}.

T. Yamakami provided another perspective on quantum problems, by
introducing a general framework for quantum inputs and quantum
operators, where quantum nondeterminism is described in a novel
way \cite{Yam02}. Furthermore, he constructed a quantum hierarchy
similar to the Meyer-Stockmeyer polynomial hierarchy, based on
two-sided bounded-error quantum computation. He also defined the
notion of quantum {\em partial decision problem} as a pair of
accepted and rejected sets of quantum states. As we will see later
these two approaches for describing quantum problems are closely
related.

In this paper, we introduce different quantum languages which
exhibit interesting relations on quantum states. In order to
analyze the complexity of these quantum languages we extend the
notion of complexity classes QMA and QCMA to quantum inputs.
Finally we discuss how to derive a classical language from a given
quantum language.

\section{Preliminaries}

We begin by defining our terms. We use Dirac's notation
$\ket{\phi}$ to describe a pure quantum state and $\varrho$ to
describe a density matrix representation for a quantum state (pure
or mixed). A pure {\em quantum string} of size $n$ is a unit-norm
vector in a Hilbert space of dimension $2^n$. For a given quantum
string $\ket{\phi}$, $\ell(\ket{\phi})$ denotes the size of
$\ket{\phi}$ (the number of qubits in $\ket{\phi}$). Following the
terminology of \cite{Yam02}, we use the notation $\Phi_n$ to
denote the set of all pure quantum strings of size $n$. Define
$\Phi_{\infty}=\bigcup_{n \geq 0} \Phi_n$, to be the set of all
finite size pure quantum strings. Since the density operator
representation for quantum strings is better suited for parts of
our discussion, we define $\Omega_n$ to be the set of all density
matrices of $n$ qubits, and $\Omega_{\infty}=\bigcup_{n \geq 0}
\Omega_n$ to be the set of all finite size density matrices.

We work within the quantum network model as a mathematical model
of quantum computation \cite{Deutsch89,Yao93}. To study the
complexity classes in the circuit model we use the concept of {\em
polynomial-time uniformly generated family}, i.e. a sequence of
quantum circuits, $\{C_{n}\}$, one for each input length $n$, that
can be efficiently generated by a Turing machine. We assume each
$C_{n}$ runs in time polynomial in $n$ and that it is composed of
gates in some reasonable, universal, finite set of quantum gates
\cite{NC00}. Furthermore, the number of gates in each $C_{n}$ is
not bigger than the length of the description of that circuit.
Therefore, the size of $C_{n}$ is polynomial in $n$. We often
identify a circuit $C$ with the unitary operator it induces. We
say that a circuit $C$ accepts a quantum input $\ket{\phi}$ with
probability $p$ if, when we run $C$ with input register in state
$\ket{\phi}$ and auxiliary registers in $\ket{0}$, we observe $1$
with probability $p$ on the output register. We denote by
$\mathrm{Prob}[C(\ket{\phi})=1]$ the acceptance probability of $C$
on input $\ket{\phi}$. It is well known that a polynomial-time
quantum Turing machine and a polynomial-time uniformly generated
family of quantum circuits are computationally equivalent.

Considering quantum states as inputs raises the following issues.
First, due to the no-cloning theorem, we cannot copy an unknown
quantum input. Therefore, to repeat the same quantum computation
over a given quantum input we assume that a quantum state is given
as a black box, from which one can prepare copies of the required
input state on request. Equivalently, we can consider that the
copies of the quantum input are given a priori. Second, since the
the space $\Phi_{\infty}$ is continuous to define the notation of
complexity classes for $\Phi_{\infty}$ we consider {\em partial
decision problem} over $\Phi_{\infty}$ \cite{Yam02}. A partial
decision problem is a pair $(A,B)$ such that $A,B \subset
\Phi_{\infty}$ and $A \cap B = \emptyset$, where $A$ indicates a
set of accepted quantum strings and $B$ indicates a set of
rejected quantum strings. The {\em legal region} of $(A,B)$ is
$A\cup B$.

Consider a {\em quantum language} $L\subset \Phi_{\infty}$. We
define the corresponding partial decision problem for an arbitrary
real number $\epsilon >0$ to be $P_{L,\epsilon}=(A,B)$, with
\begin{eqnarray*}
A&=&L \\
B&=&\{ \ket{\psi}\in\Phi_{\infty}\,\mid\, \forall \ket{\phi} \in L
\,:\,
\parallel \ket{\psi} - \ket{\phi}
\parallel \geq \epsilon\} \, ,
\end{eqnarray*}
where extra $\ket{0}$'s are added to make $ \parallel . \parallel$
meaningful. In other words there is an {\em illegal region} where
$P_{L,\epsilon}$ cannot decide, and the size of this region is
bounded by $\epsilon$. Note that deciding $P_{L,\epsilon}$ is
equivalent to test the global property $P$ which defines the
quantum states in $L$, since a given quantum state $\ket{\phi}$
either satisfies the property $P$ (belongs to $A$) or it is far
from all the quantum states satisfying $P$ (belongs to $B$).

Now, in this general framework a complexity class denoted by
${}^*{\cal C}$ is a collection of partial decision problems.
Yamakami described these complexity classes in terms of a well
formed quantum Turing machine with access to polynomial number of
copies of quantum states \cite{Yam02}. Equivalently, we work
within the uniform circuit family where polynomial number of
copies of the input state are given a priori.

\begin{definition}\label{d_bqp}
A partial decision problem $(A,B)$ is in ${}^*\mathrm{BQP}$ if
there exists a polynomial-time uniformly generated family of
quantum networks $\{C_n\}$ such that for every $\ket{\phi} \in
\Phi_{\infty}$ there exists a polynomial function $q$ and a unique
circuit $C_m$ with $ m=\mathrm{poly}(\ell(\ket{\phi}))$ where:
\begin{itemize}
\item [i)] if $\ket{\phi} \in A$ then
$\mathrm{Prob}[C_m(\ket{\phi}^{\otimes q(\ell(\ket{\phi}))})=1]
\geq 2/3$\, ,

\item [ii)] if $\ket{\phi} \in B$ then
$\mathrm{Prob}[C_m(\ket{\phi}^{\otimes q(\ell(\ket{\phi}))})=1]
\leq 1/3$\, \footnote{We can replace $1/3$ by any arbitrary small
number $1/\mathrm{poly}(\ell(\ket{\phi}))$.} .
\end{itemize}
\end{definition}
The other complexity classes that we will refer to are
$\mathrm{QMA}$ and $\mathrm{QCMA}$, introduced by Knill, Kitaev
and Watrous \cite{Knill96,Kitaev99,Watr00}. There are several
known $\mathrm{QMA}$ and $\mathrm{QCMA}$ languages
\cite{Kitaev99,Watr00,AR03,KR03,JWB03,WJB03}. Informally speaking
the complexity class $\mathrm{QMA}$ ($\mathrm{QCMA}$) is the class
of classical decision problems for which a YES answer can be
verified by a quantum computer with access to a quantum
(classical) proof. In the next section we introduce several
partial decision problems in ${}^*\mathrm{QMA}$ and
${}^*\mathrm{QCMA}$.

\begin{definition}\label{d_qma}
A partial decision problem $(A,B)$ is in ${}^*\mathrm{QMA}$ if
there exists a polynomial-time uniformly generated family of
quantum networks $\{C_n\}$ such that for every $\ket{\phi} \in
\Phi_{\infty}$ there exists a polynomial function $q$ and a unique
circuit $C_m$ with $ m=\mathrm{poly}(\ell(\ket{\phi}))$ where:
\begin{itemize}
\item [i)] if $\ket{\phi} \in A \;\;\;$ then $\;\;\;\exists
\ket{\xi} \in \Phi_{\infty}$ with
$\ell(\ket{\xi})=\mathrm{poly}(\ell(\ket{\phi})) \;: \\
\;\;\mathrm{Prob}[C_m(\ket{\phi}^{\otimes
q(\ell(\ket{\phi}))}\ket{\xi}^{\otimes q(\ell(\ket{\phi}))})=1]
\geq 2/3$,

\item [ii)] if $\ket{\phi} \in B\;\;\;$ then $\;\;\;\forall
\ket{\xi} \in \Phi_{\infty}$ with
$\ell(\ket{\xi})=\mathrm{poly}(\ell(\ket{\phi})) \; : \\
\mathrm{Prob}[C_m(\ket{\phi}^{\otimes
q(\ell(\ket{\phi}))}\ket{\xi}^{\otimes q(\ell(\ket{\phi}))})=1]
\leq 1/3$ \, .
\end{itemize}
\end{definition}
The definition of the complexity class ${}^*\mathrm{QCMA}$ is
similar to the above, except that instead of $\ket{\xi} \in
\Phi_{\infty}$, we consider a classical state $x \in \Sigma^*$ as
the proof. All the above definitions can naturally be extended to
$\Omega_{\infty}$.

It is important to note that in this paper we consider a quantum
state to represent only the data. The case where a quantum state
describe a quantum program has been studied in \cite{NC97}, where
authors argued that to represent $N$ distinguishable quantum
programs (unitary operators), $N$ orthogonal states are required.
Since the number of possible unitary operations on $m$ qubits is
infinite, it follows that a universal quantum machine with quantum
input as program would require an infinite number of qubits and
thus no such machine exists.

The final concept we introduce is a simple quantum network that
can be used as a basic building block for direct quantum
estimations of both linear and non-linear functionals of any
quantum state $\varrho$~\cite{EMOHHK02,MHOKE02}. The network can
be realized as multiparticle interferometry and it provides a
direct estimation of the overlap of any two unknown quantum states
$\varrho_a$ and $\varrho_b$, i.e. $\tr\varrho_a\varrho_b$.

\begin{figure}
\begin{center}
\epsfig{figure=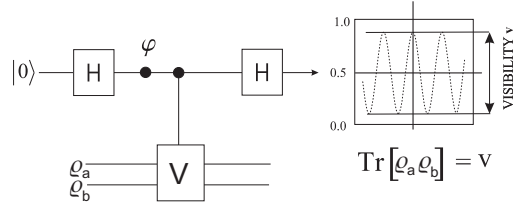,width=0.55\textwidth}
\end{center}
\caption{A quantum network for direct estimations of both linear
and non-linear functions of state. The probability of finding the
control (top line) qubit in state $\ket{0}$ at the output depends
on the overlap of the two target states (two bottom lines). Thus
estimation of this probability leads directly to an estimation of
$\tr \varrho_a\varrho_b=\vis=2\,P_0-1$.} \label{f_En}
\end{figure}

In order to explain how the network works, let us start with a
general observation related to modifications of visibility in
interferometry. Consider a typical interferometric set-up for a
single qubit: Hadamard gate, phase shift gate $\varphi$, Hadamard
gate, followed by a measurement in the computational basis (Figure
\ref{f_En}). We modify the interferometer by inserting a
controlled-$V$ operation between the Hadamard gates, with its
control on the single qubit and with $V$ acting on two quantum
systems described by $\varrho_a$ and $\varrho_b$ respectively. The
operator $V$ is the swap operator, defined as
$V\ket{\alpha}_{A}\ket{\beta}_{B}=\ket{\beta}_{A}\ket{\alpha}_{B}$,
for any pure states $\ket{\alpha}_{A}$ and $\ket{\beta}_{B}$. The
action of the controlled-$V$ on $\varrho_a\otimes\varrho_b$
modifies the interference pattern by the factor
\begin{equation*}
\vis = \tr V\left(\varrho _{a}\otimes\varrho
_{b}\right)=\tr\varrho_{a}\varrho_{b}, \label{eqvisi}
\end{equation*}
where $\vis$ is the new visibility. The observed modification of
the visibility gives an estimate of $\tr (\varrho_a \varrho_b)$,
i.e. the overlap between states $\varrho_a$ and $\varrho_b$. The
probability of finding the control qubit in state $\ket{0}$ at the
output, $P_0$, is related to the visibility by $ \vis=2\,P_0-1$.
The above network is one of the main ingredients for our
discussion in the next section, we redefine it as follows:
\begin{definition}
Let $n$ be an integer number. The following quantum network with
$2n+1$ qubits is called {\em estimation} network and is denoted by
$E_n :$
\begin{eqnarray*}
(H^1\otimes I^n\otimes I^n) \circ (\mbox{ctrl-}V) \circ
(H^1\otimes I^n\otimes I^n) \, .
\end{eqnarray*}
\end{definition}

\section{Quantum Languages}

We start by introducing a simple language in ${}^*\mathrm{BQP}$
and we build up towards more interesting languages in
${}^*\mathrm{QCMA}$ and ${}^*\mathrm{QMA}$. In what follows we say
that a language $L$ in $\Phi_{\infty}$ or $\Omega_{\infty}$
belongs to a complexity class ${}^*C$ iff there exists a small
real number $\epsilon$ such that the corresponding partial
decision problem $P_{L,\epsilon}$ lies in ${}^*C$.

\begin{figure}
\begin{center}
\epsfig{figure=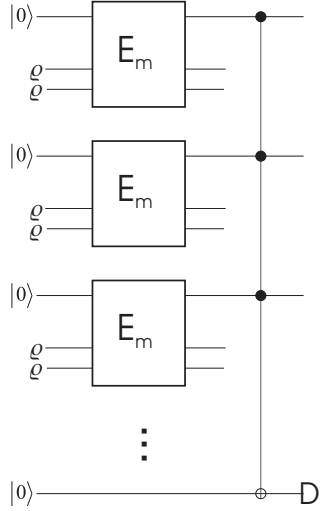,width=0.35\textwidth}
\end{center}
\caption{A quantum network for accepting Language $L_1$. If
$\varrho$ (the state of the first $m$ qubits of $\ket{\phi}$) is
pure then the last qubit will be in state $\ket{1}$ with
probability one. On the other hand if $\varrho$ is mixed the
probability of measuring $1$ in the last qubit decreases
exponentially with the number of runs.} \label{f_L1}
\end{figure}

\begin{example}
Let $f:\mathbb{N}\rightarrow\mathbb{N}$ be a function such that
for all $n$ we have $f(n)\leq n$. The quantum language $L_1$
defined below belongs to ${}^*\mathrm{BQP}:$
\begin{equation*}
L_1=\{\;\ket{\phi} \in \Phi_{\infty} \;:\; \mbox{The state of the
first $f(\ell(\ket{\phi}))$ qubits of $\ket{\phi}$ is pure}\} \, .
\end{equation*}
\end{example}
\begin{proof}
Let $\varrho$ be the state of the first $m=f(\ell(\ket{\phi}))$
qubits of $\ket{\phi}$ (this can be prepared by tracing out the
rest of the qubits in  $\ket{\phi}$) and apply $E_m$ to
$\ket{0}\otimes\varrho\otimes\varrho$. The probability of
observing $0$ in the first register is :
\begin{equation*}
P_0=\frac{\tr(\varrho^2)+1}{2} \, .
\end{equation*}
If $\varrho$ is pure, then $P_0=1$. If $\varrho$ is mixed, then
$1/2<P_0<1$. Hence, in order to check that $\varrho$ is indeed
pure we need to run $E_m$ for a polynomial number of times
$M=\mathrm{poly}(\ell(\ket{\phi}))$ and measure the state of the
control qubit. In this case $P_0=(\frac{\tr(\varrho^2)+1}{2})^M$,
which will be equal to $1$ if $\varrho$ is pure or tend
exponentially to $0$ if $\varrho$ is mixed. The probability of
accepting $\varrho$ as pure when it is in fact mixed is thus
exponentially small on the number of runs of $E_m$. Therefore the
following polynomial-time uniformly generated family of quantum
circuits $\{ C_n \}$ satisfies the condition of the definition
\ref{d_bqp}, and will do the job (Figure \ref{f_L1}):
\begin{eqnarray*}
C_m = T_{M,1} \circ (E_m\otimes E_m\otimes\cdots\otimes E_m) \, .
\end{eqnarray*}
where $M=\mathrm{poly}(\ell(\ket{\phi}))$ and $T_{M,1}$ is a
Toffoli type gate which flips the last qubit if all the first $M$
qubits are equal to $1$. Note that the the number of qubits of
each $C_m$ is polynomial in $\ell(\ket{\phi})$.
\end{proof}

The next example is an extension of $L_1$ and belongs to
${}^*\mathrm{QCMA}$. We can view Definition \ref{d_qma} for
accepting a language in ${}^*\mathrm{QCMA}$ or ${}^*\mathrm{QMA}$
as an interactive protocol consisting of two parties, often called
Merlin (with unlimited computational power) and Arthur (with
quantum polynomial-time power). Merlin is trying to persuade
Arthur that a quantum state $\varrho$ in a given language $L$
satisfies a given property. To this end, he sends Arthur a
polynomial-size classical or quantum state as the proof. Each
party has access to polynomial number of copies of $\varrho$.
Therefore, a protocol to accept the language will be a
polynomial-time uniformly generated family of quantum networks
with classical or quantum inputs given by Merlin and possibly
polynomial number of copies of state $\varrho$.

\begin{example}
The quantum language $L_2$ defined below belongs to
${}^*\mathrm{QMCA}:$
\begin{equation*}
L_2=\{\;\ket{\phi} \in \Phi_{\infty} \;:\; \mbox{$\ket{\phi}$ is
separable with respect to two disjoint subsets of qubits}\} \, .
\end{equation*}
In other words every $\ket{\phi} \in L_2$ can be written as
$\ket{\phi_1}\otimes\ket{\phi_2}$ where
$\ell(\ket{\phi_1})+\ell(\ket{\phi_2})=\ell(\ket{\phi})$.
\end{example}
\begin{proof}
Since we assume $\ket{\phi}$ to be pure, it follows that
$\ket{\phi_1},\ket{\phi_2}$ will be pure as well:
\begin{eqnarray*}
\tr(\ket{\phi}\bra{\phi}^2) &=&
\tr(\ket{\phi_1}\otimes\ket{\phi_2}\bra{\phi_1}\otimes\bra{\phi_2}^2)
\\ &=&
\tr(\ket{\phi_1}\bra{\phi_1}^2)\tr(\ket{\phi_2}\bra{\phi_2}^2)=1
\end{eqnarray*}
therefore:
\begin{eqnarray*}
\tr(\ket{\phi_1}\bra{\phi_1}^2)=\tr(\ket{\phi_2}\bra{\phi_2}^2)=1
\, .
\end{eqnarray*}

The protocol for accepting $L_2$ uses the network family $\{C_n\}$
of the previous example. During the protocol, for $\ket{\phi}\in
L_2$ Merlin will send a classical binary string of the size
$l(\ket{\phi})$, called {\em subset string}, where each $1$ at
position $i$ indicates that the $i$th qubit in $\ket{\phi}$
appears in subset ($\ket{\phi_1}$). Given a subset string $S$ and
the corresponding quantum state $\ket{\phi}$, Arthur apply the
simple network of Figure \ref{f_L2} to prepare the corresponding
subset state and checks its purity with the proper network $C_m$
of the previous example. If Merlin attempts to cheat by sending a
false partition the probability of obtaining a $1$ will decrease
exponentially with the number of runs.

\begin{figure}
\begin{center}
\epsfig{figure=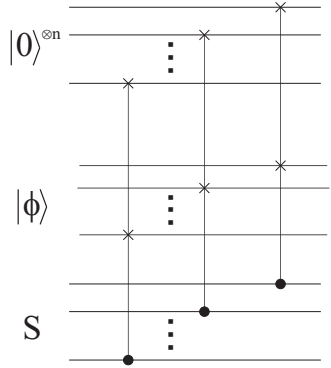,width=0.35\textwidth}
\end{center}
\caption{A quantum network to prepare the corresponding subset
state of the state $\ket{\phi}$. In the given classical string
$S$, each $1$ at the position $i$ indicates that the $i$th qubit
in $\ket{\phi}$ appears in the subset state $\ket{\phi_1}$.}
\label{f_L2}
\end{figure}

Note that any pure state $\ket{\phi}$, entangled with respect to
two disjoint subsets of qubits is of the form
\begin{eqnarray*}
\ket{\phi}=\sum_{i=1}^N c_i \ket{\phi_1^i}\otimes\ket{\phi_2^i},
\end{eqnarray*}
where $\sum_i |c_i|^2=1$, $\ket{\phi_1^i}\otimes\ket{\phi_2^i}$
are an orthonormal set of states and at least two $c_i\neq 0$. On
the other hand, a pure separable state is simply
$\ket{\varphi_1}\otimes\ket{\varphi_2}$. Therefore, $\ket{\phi}$
is almost separable if there exists a small real number
$\epsilon'$ and a $1\leq j\leq N$ such that $|c_j|=1-\epsilon'$.
If $\epsilon'\leq \epsilon/2$, where $\epsilon$ is the parameter
defining the illegal region of $L_2$, $\ket{\phi}$ is undecidable.

\end{proof}

\begin{example}
The quantum language $L_3$ defined below belongs to
${}^*\mathrm{QMA}:$
\begin{equation*}
L_3=\{\;\varrho \in \Omega_{\infty} \;:\; \varrho \;\mbox{is an
entangled state}\}.
\end{equation*}
\end{example}
\begin{proof}
A quantum state is said to be entangled if it cannot be written in
the form
\begin{eqnarray*}
\varrho_{123...N}= \sum_{\ell} C_\ell \varrho^{\ell}_1 \otimes
\varrho^{\ell}_2 \otimes \varrho^{\ell}_3 \otimes \ldots \otimes
\varrho^{\ell}_N, \label{rho}
\end{eqnarray*}
where $\varrho^{\ell}_j$ is the state of subsystem $j$, and
$\sum_{\ell} C_\ell=1$. Checking that a generic
$\varrho_{123...N}$ is separable is a hard problem. 
However, it is possible to construct entanglement witnesses that
detect the entanglement in specific entangled states, provided
that the state is known. An entanglement witness $W$
~\cite{Terhal01,LKCH00} is an operator with non-negative
expectation value on all separable states, and for which there
exists an entangled state such that the expectation value of the
witness on that state is negative. Therefore, if Merlin wants to
persuade Arthur that a given state $\varrho$ is entangled, it is
sufficient for him to send Arthur the respective entanglement
witness.

Merlin cannot send the operator $W$ directly as a physical state
because even though $W$ is an operator in the density operators'
Hilbert space, it will not be a valid state in general. So, during
the protocol for $\ket{\phi}\in L_3$, Merlin will send $W$ as a
set of density operators $\varrho_1,...,\varrho_k$ (each of the
dimension of $\ket{\phi}$) and a classical string of $k$ real
numbers $c_i$, such that $W=\sum_i c_i \varrho_i$. Note that based
on the construction of a generic $W$ in \cite{LKCH00}, the number
of $\varrho_i$'s is polynomial in $\ell(\ket{\varrho})$ and
$c_i$'s are polynomial computable real numbers.

Now, Arthur uses $\varrho_i$ and $\ket{\phi}$ as inputs to the
proper $C_m$ and computes $\tr(\varrho_i \ket{\phi})$, for each
$i$. Using these expectation values and numbers $c_i$, Arthur can
estimate
\begin{equation*}
\tr(\sum_i c_i \varrho_i \ket{\phi})=\tr(W \ket{\phi}) \, .
\end{equation*}
If he obtains a negative value, he knows that $\ket{\phi}$ is
entangled. Also, in order to check that Merlin did send him an
entanglement witness, he can prepare the basis of separable states
(with polynomial-size) and check that the expectation value on
these states is non-negative.

\end{proof}

To introduce the following language we need few definitions from
\cite{KNV02}. Kashefi et al. studied the relation between
preparing a set of quantum states and constructing the reflection
operators about those states. We begin by the following natural
definition of the ``easy'' states and operators:

\begin{definition}
A unitary operator $U$ on $n$ qubits is {\em polynomial-time
computable} (easy), if there exists a network approximately
implementing $U$ with polynomial-size in $n$. An $n$-qubit state
$\ket{\phi}$ is defined to be {\it polynomial-time preparable}
(easy), if there exists an easy operator $U$ on ${\rm poly}(n)$
qubits such that $U|0\rangle=\ket{\phi}$.
\end{definition}
It is well-known that if a state $\ket{\phi}$ is easy, then the
reflection operator about that state, $2\ket{\phi}\bra{\phi}-I$,
is easy (Problem $6.2(1)$ in \cite{NC00}). The inverse statement
is called the {\em Reflection Assumption}:  ``if the reflection
about a state is easy, the state itself is easy'' and it is known
that:

\begin{lemma} \cite{KNV02}
If there exists a quantum one-way function, then exists a
counter-example to the Reflection Assumption.
\end{lemma}

The next quantum language is concerned with reflection operators:

\begin{example}
The quantum language $L_4$ defined below belongs to
${}^*\mathrm{QCMA}$:
\begin{equation*}
L_4=\{\;\ket{\phi} \in \Phi_{\infty} \;:\; \mbox{The operator
$\;\;2\ket{\phi}\bra{\phi}-I\;\;$ is polynomial-time
computable}\}\, .
\end{equation*}
\end{example}
\begin{proof}

During the protocol, for $\ket{\phi}\in L_4$ Merlin will send a
classical description of the polynomial-size quantum network
implementing the reflection operator
$R_{\phi}=2\ket{\phi}\bra{\phi}-I$. Arthur prepares an arbitrary
state $\ket{\xi_i}$ (unknown to Merlin), which can always be
written as:
\begin{equation*}
\ket{\xi_i}=\alpha \ket{\phi}+\beta\ket{\phi^{\perp}}
\end{equation*}
and applies $R_{\phi}$ to a copy of $\ket{\xi}$, obtaining the
output state
\begin{equation*}
\ket{\xi_o}=\alpha\ket{\phi} -\beta\ket{\phi^{\perp}}\, .
\end{equation*}
Then he uses the network $E_m$ with $m=\ell{\ket{\phi}}$ to
compute the following values unknown to Merlin:
\begin{eqnarray}
\inner{\xi_i}{\phi} &=& \mid\alpha\mid^2 \, , \\
\inner{\xi_o}{\phi} &=& \mid\alpha\mid^2 \, , \\
\inner{\xi_o}{\xi_i} &=& \left |\mid\alpha\mid^2 -
\mid\beta\mid^2\right | \, ,
\end{eqnarray}
and he repeats this procedure for
$M=\mathrm{poly}(\ell(\ket{\phi}))$ different $\ket{\xi_i}$.
Arthur will accept $\ket{\phi}$ iff at each run of the above
procedure the value of $\alpha$ obtained from Equation $1$
satisfies Equation $2$ and $3$.

Now assume that $\ket{\phi}$ is far from all elements of $L_4$,
i.e. the reflection operator about $\ket{\phi}$ is not easy, and
that Merlin attempts to cheat by sending the description of a
polynomial-size network $N$, where $N\neq R_{\phi}$. Following the
above strategy, when Arthur applies $N$ to
$\ket{\xi_i}=\alpha\ket{\phi}+\beta\ket{\phi^{\perp}}$ he will
obtain a state of the form
$\ket{\xi_o}=\alpha^{\prime}\ket{\phi}+\beta^{\prime}\ket{\phi^{\perp}}$.
If now he computes the values for the Equation $1$, $2$ and $3$,
he will get
\begin{eqnarray*}
\inner{\xi_i}{\phi} &=& \mid\alpha\mid^2 \, , \\
\inner{\xi_o}{\phi} &=& \mid\alpha^{\prime}\mid^2 \, , \\
\inner{\xi_o}{\xi_i} &=& \left
|\alpha^*\alpha^{\prime}+\beta^*\beta^{\prime}\right | \, .
\end{eqnarray*}
If $\mid\alpha\mid^2\neq\mid\alpha^{\prime}\mid^2$ Arthur will
detect the cheating. If on the other hand
$\mid\alpha\mid^2=\mid\alpha^{\prime}\mid^2$, which implies that
$\mid\beta\mid^2=\mid\beta^{\prime}\mid^2$, we have that
$\inner{\xi_o}{\xi_i} = \left |\mid\alpha\mid^2+ e^{i
\theta}\mid\beta\mid^2\right |$, where $\theta$ is the relative
phase between $\beta$ and $\beta^{\prime}$. Whenever $\theta \neq
\pi$, Equation $3$ will not be satisfied and Arthur will detect
the cheating.
\end{proof}

Another interesting language in close relation to $L_4$ is defined
below. First we define the notion of a {\em polynomial-time
checkable} state \cite{ZNote}.

\begin{definition}
We define a state $\ket{\phi}$ to be {\em efficiently checkable}
if there exits a polynomial-size quantum network implementing the
following checking operator:
\begin{eqnarray*}
C_{\phi}\ket{\phi}\ket{0} &=& \ket{\phi}\ket{0} \\
C_{\phi}\ket{\psi}\ket{0} &=&
\alpha\ket{\omega}\ket{0}+\beta\ket{\psi}\ket{1} \;\; \mbox
{where} \\ \forall \ket{\psi}\perp\ket{\phi} \;&:&\; \alpha = 0 \,
.
\end{eqnarray*}
\end{definition}

\begin{example}
The quantum language $L_5$ defined below belongs to
${}^*\mathrm{QCMA}$:
\begin{equation*}
L_5=\{\;\ket{\phi} \in \Phi_{\infty} \;:\; \ket{\phi} \; \mbox{is
efficiently checkable}\}.
\end{equation*}
\end{example}
\begin{proof}
The next lemma shows that $L_4=L_5$ which implies $L_5 \in
{}^*\mathrm{QCMA}$.
\end{proof}
\begin{lemma}
The quantum languages $L_4$ and $L_5$ are equal.
\end{lemma}
\begin{proof}
Denote by ctrl-$R_{\phi}$ the controlled reflection operator which
reflects the state of the first register about $\ket{\phi}$ iff
the last qubit (the control qubit) is equal to $1$. We show for
any state $\ket{\phi}$:
\begin{equation*}
C_{\phi}=I\otimes H \;\circ\; \mbox{ctrl-}R_{\phi} \;\circ\;
I\otimes H \, ,
\end{equation*}
and therefore $L_4=L_5$. It is easy to check:
\begin{eqnarray*}
\ket{\psi}\ket{0} &\stackrel{I\otimes H}{\longrightarrow}&
\frac{1}{\sqrt{2}}\ket{\psi}(\ket{0}+\ket{1})\\
&\stackrel {\mbox{ctrl-}R_{\phi}}{\longrightarrow}&
\frac{1}{\sqrt{2}}\{\ket{\psi}\ket{0}+(2\inner{\phi}{\psi}\ket{\phi}-\ket{\psi}\ket{1})\}\\
&\stackrel {I\otimes H}{\longrightarrow}&
\frac{1}{2}\{2\inner{\phi}{\psi}\ket{\phi})\ket{0}+2(\ket{\psi}-\inner{\phi}{\psi}\ket{\phi})\ket{1}\}
\, .
\end{eqnarray*}
If $\ket{\psi}\perp \ket{\phi}$ the final state of the above
computation is $\ket{\psi}\ket{1}$ as required.
\end{proof}

\section{Discussion}

Following the work of Yamakami \cite{Yam02}, we have considered a
general framework for quantum machines with quantum states as
input. We introduced some quantum languages in this paradigm and
showed the corresponding partial decision problems belong to
complexity classes ${}^*\mathrm{BQP}$, ${}^*\mathrm{QCMA}$ and
${}^*\mathrm{QMA}$. These quantum languages can also be viewed as
quantum property testing of a set of quantum states.

This investigation of quantum properties (quantum languages) is
useful for defining new classical languages within the framework
of quantum information theory. For instance, if we consider the
subset of quantum states that can be prepared in polynomial-time,
e.g. with a polynomial-size quantum circuit, we can derive a
classical language from the given quantum language. However, it is
not clear how to extend a given classical language to its quantum
counterpart. As an example of this derivation consider the
classical analogue of language $L_3$:
\begin{eqnarray*}
L^{\prime}_3= \{ x \in \Sigma^* \;&:&\; x \; \mbox {describes a
polynomial-size quantum network} \; U \\ & & \mbox {and} \;
U\ket{0} \; \mbox{is an entangled state} \} \, ,
\end{eqnarray*}
which belongs to $\mathrm{QMA}$.

Recently, few complete languages for $\mathrm{QCMA}$ and
$\mathrm{QMA}$ have been introduced. Finding complete languages
for ${}^*\mathrm{QCMA}$ and ${}^*\mathrm{QMA}$ would also be very
interesting, but it is so far an open problem.

\section*{Acknowledgements}
We are grateful to Harumichi Nishimura for useful comments and
suggestions. EK thanks Hirotada Kobayashi, Frederic Magniez and
Keiji Matsumoto for insightful discussions on the topic of quantum
languages. CMA is supported by the Funda{\c c}{\~a}o para a
Ci{\^e}ncia e Tecnologia (Portugal).


\begin{thebibliography}{10}
\bibitem{ZNote} The notion of efficiently checkable state and its
relation to language $L_4$ has been mentioned to us by Frederic
Magniez.
\end{thebibliography}

\end{document}